\documentclass[a4paper]{jpconf}
\usepackage{graphicx}
\usepackage[numbers,sort&compress]{natbib}
\usepackage{color} 
 \usepackage{soul} 
\begin{document}
\title{Understanding globular cluster abundances through nuclear reactions}

\author{P Adsley$^{1,2,3,4}$, M Williams$^5$, D S Harrouz$^6$, D P Carrasco-Rojas$^{2,7}$, N de Séréville$^6$, F Hammache$^6$, R Longland$^{8,9}$, B Bastin$^{10}$, B Davids$^{5,11}$, T Faestermann$^{12}$, C Foug\`eres$^{10}$, U Greife$^{13}$, R Hertenberger$^{14}$, D Hutcheon$^5$, M La Cognata$^{15}$, AM Laird$^{16}$, L Lamia$^{15,17}$, A Lennarz$^5$, A Meyer$^6$, F d'Oliveira Santos$^{10}$, S Palmerini$^{18,19}$, A Psaltis$^{20}$, R G Pizzone$^{15}$, S Romano$^{15,17,21}$, C Ruiz$^{5,22}$, A Tumino$^{15,23}$, H-F Wirth$^{14}$}

\address{$^1$ Department of Physics and Astronomy, Texas A\&M University, College Station, 77843-4242, Texas, USA}
\address{$^2$ Cyclotron Institute, Texas A\&M University, College Station, 77843-3366, Texas, USA}
\address{$^3$ School of Physics, University of the Witwatersrand, Johannesburg 2050, South Africa}
\address{$^4$ iThemba Laboratory for Accelerator Based Sciences, Somerset West 7129, South Africa}
\address{$^5$ TRIUMF, Vancouver, BC V6T 2A3, Canada}
\address{$^6$ Universit\'e Paris-Saclay, CNRS/IN2P3, IJCLab, 91405 Orsay, France}
\address{$^7$ Department of Physics, University of Texas at El Paso, El Paso, TX 79968-0515, USA}
\address{$^8$ North Carolina State University, Raleigh, North Carolina 27695, USA}
\address{$^{9}$ Triangle Universities Nuclear Laboratory, Durham, North Carolina 27708, USA}
\address{$^{10}$ GANIL, CEA/DRF-CNRS/IN2P3, Bvd Henri Becquerel, 14076 Caen, France}
\address{$^{11}$ Department of Physics, Simon Fraser University, 8888 University Drive, Burnaby, British Columbia V5A 1S6, Canada}
\address{$^{12}$ Physik Department E12, Technische Universität München, D-85748 Garching, Germany}
\address{$^{13}$ Department of Physics, Colorado School of Mines, Golden, Colorado 80401, USA}
\address{$^{14}$ Fakultät für Physik, Ludwig-Maximilians-Universität München, D-85748 Garching, Germany}
\address{$^{15}$ Laboratori Nazionali del Sud–Istituto Nazionale di Fisica Nucleare, Via Santa Sofia 62, 95123 Catania, Italy}
\address{$^{16}$ Department of Physics, University of York, Heslington, York, YO10 5DD, United Kingdom}
\address{$^{17}$ Dipartimento di Fisica e Astronomia E. Majorana, Università di Catania, 95123 Catania, Italy}
\address{$^{18}$ Dipartimento di Fisica e Geologia, Università degli Studi di Perugia, 06123 Perugia, Italy}
\address{$^{19}$ Istituto Nazionale di Fisica Nucleare, Sezione di Perugia, 06123 Perugia, Italy}
\address{$^{20}$ Institut f\"{u}r Kernphysik, Technische Universit\"{a}t Darmstadt, Schlossgartenstr. 2, Darmstadt 64289, Germany}
\address{$^{21}$ Centro Siciliano di Fisica Nucleare e Struttura della Materia (CSFNSM), 95123 Catania, Italy}
\address{$^{22}$ Department of Physics \& Astronomy, University of Victoria, Victoria, BC V8W 2Y2, Canada}
\address{$^{23}$ Facoltà di Ingegneria e Architettura, Università degli Studi di Enna, 94100 Enna, Italy}

\ead{padsley@tamu.edu}

\begin{abstract}
Globular clusters contain multiple stellar populations, with some previous generation of stars polluting the current stars with heavier elements. Understanding the history of globular clusters is helpful in understanding how galaxies merged and evolved and therefore constraining the site or sites of this historic pollution is a priority. The acceptable temperature and density conditions of these polluting sites depend on critical reaction rates. In this paper, three experimental studies helping to constrain astrophysically important reaction rates are briefly discussed.
\end{abstract}

\section{Astrophysical Motivation and Nuclear-Data Needs}

Globular clusters were once thought to be composed of single populations of ancient stars. In the last few decades, it has become apparent that they instead are host to multiple stellar populations \cite{doi:10.1146/annurev-astro-081817-051839}. One piece of evidence for these multiple populations is the observation of abundance anomalies within the current generation of stars which cannot originate with the current stars due to their limited temperatures. Instead, some previous polluting site or sites has caused these abundance anomalies. However, the nature of these now extinct sites is unclear: uncertainties in thermonuclear reaction rates in stars lead to associated uncertainties over the temperature and density conditions in which the abundance anomalies could have originated. Better constraints on the reaction rates lead to better constraints on the astrophysical site or sites.

In order to determine the astrophysical reaction rates of interest, the cross section at each energy, $\sigma(E)$ must be folded with the Maxwell-Boltzmann distribution describing the relative energy distribution of all of the ions within the hot stellar plasma. For many reactions in the mass region relevant for nucleosynthesis in globular clusters, especially for radiative-capture reactions, the contribution of a resonance to the reaction rate can be simplified by using the narrow-resonance approximation. This is valid when the width of the resonance is so narrow that the width appears constant over the region of interest. In this case, the contribution of narrow resonances to the reaction rate at a temperature $T_9$ in GK can be given as:
\begin{equation}
    N_A \langle \sigma v \rangle = \frac{1.54 \times 10^{11}}{\left(\mu T\right)^{3/2}}\sum_i (\omega\gamma)_i e^{-11.605E_{r,i}/T_9}\ \mathrm{cm}^{3}\ \mathrm{s}^{-1}\  \mathrm{mol}^{-1}
    \label{eq:narrowres}
\end{equation}
where the spin-multiplicity factor in terms of the spins of the reactants $j_{1,2}$ and of the resonance $J$ is $\omega = \frac{2J+1}{(2j_1+1)(2j_2+1)}$, $\gamma = \frac{\Gamma_p\Gamma_\gamma}{\Gamma_p + \Gamma_\gamma}$ is given in terms of the proton width $\Gamma_p$ and the $\gamma$-ray width $\Gamma_\gamma$ for a ($p,\gamma$) radiative-capture reaction, $E_r$ is the resonance energy in MeV, and the sum across $i$ is for all of the narrow resonances which can contribute to the reaction rate. The product $\omega\gamma$ is known as the resonance strength.

From Equation \ref{eq:narrowres}, it is clear that the following pieces of information are required for the calculation of the reaction rate: the existence of a state, the resonance energy $E_r$, the spin, $J$ and the resonance strength or partial widths. In this paper, three different experiments probing different ways to access these vital pieces of information are discussed, showing the importance of combining direct and indirect measurements for discoveries in nuclear astrophysics.

\section{$^{22}$Ne($p,\gamma$)$^{23}$Na and the $^{23}$Na($p,p^\prime$)$^{23}$Na$^*$ reaction}

The $^{22}$Ne($p,\gamma$)$^{23}$Na reaction is critical to understanding the sodium-oxygen anticorrelation observed in many globular clusters. This reaction has been the focus of many direct measurements, including using the DRAGON recoil separator at TRIUMF \cite{LENNARZ2020135539,PhysRevC.102.035801,HUTCHEON2003190}, at the Laboratory for Experimental Nuclear Astrophysics (LUNA) \cite{PhysRevC.95.015806} and at the Laboratory for Underground NuclearAstrophysics (LUNA) in Gran Sasso, Italy \cite{PhysRevC.94.055804,PhysRevLett.121.172701}. Major sources to the uncertainty in the reaction rate have originated from two tentative resonances located at $E_r = 65$ and $100$ keV ($E_\mathrm{x} = 8862$ and $8894$ keV, respectively). The higher of these two resonances was ruled out as unimportant through one LUNA experiment \cite{PhysRevLett.121.172701} but the lower $E_r = 65$ keV resonance may still increase the reaction rate considerably. The evidence for the existence of this state is unclear. One $^{22}$Ne($^3$He$,d$)$^{23}$Na experiment reported tentative discovery of the states \cite{PhysRevC.4.2030} but subsequent measurements of the same reaction \cite{PhysRevC.65.015801} have failed to observe them. 

The fundamental question that must be resolved regarding these resonances is therefore \textquoteleft do these resonances actually exist?\textquoteright\ Since the direct measurements at LUNA have been unable to observe any signal from these resonances, and the $^{22}$Ne($^3$He$,d$)$^{23}$Na experiment of Hale {\it et al.} also failed to observe this state, there is a paucity of evidence for existence. To try to answer the question of whether the states which could give rise to these resonances exist, we have used the proton inelastic-scattering reaction at low energies to populate states in $^{23}$Na. This reaction was performed with a $E_p = 14$-MeV beam. This means that the states populated in $^{23}$Na are not being populated in resonance scattering but in some other, rather non-selective, reaction mechanism which appears to be compound in nature. This reaction has been used to study a number of different isotopes already, such as $^{31}$P, $^{28}$Si, $^{27}$Al, $^{24,26}$Mg, $^{23}$Na, and $^{19}$F \cite{PhysRevC.103.035804,PhysRevC.97.045807,PhysRevC.102.015801,MOSS1969440,MOSS1976413,MOSS1976429,PhysRevC.89.065805}.

\begin{figure}
    \centering
    \begin{minipage}[b]{3.5in}
    \includegraphics[width=\textwidth]{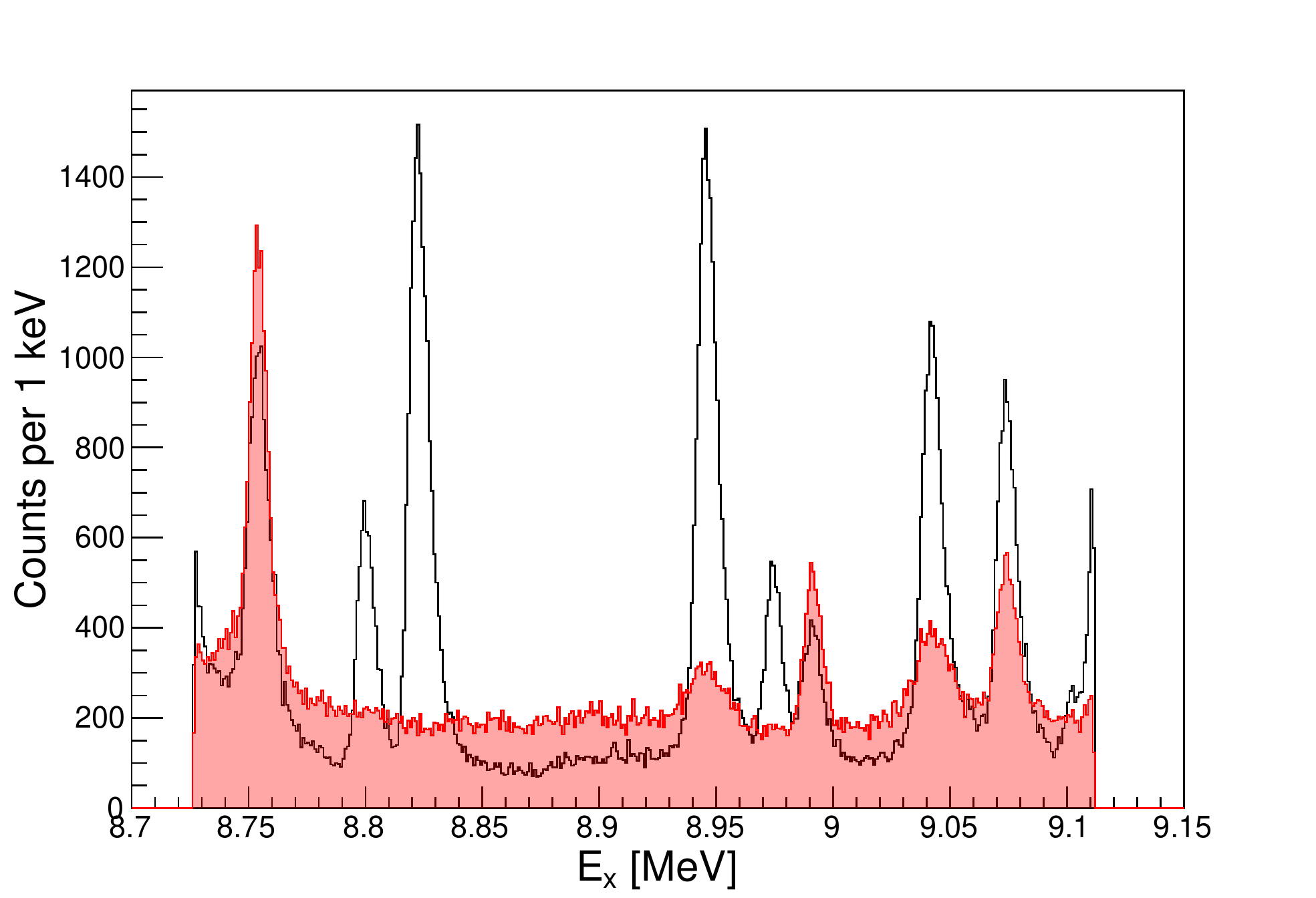}
    \end{minipage}
    \hspace{0.2cm}
    \begin{minipage}[b]{2.5in}
    \caption{The excitation-energy spectrum from the $^{23}$Na($p,p^\prime$)$^{23}$Na$^*$ reaction measured at $\theta_\mathrm{Q3D} = 70$ degrees. The black hollow spectrum is that resulting from the NaF target and the filled red spectrum is from a LiF target. }
    \label{fig:na23pp_spectrum}
    \end{minipage}
\end{figure}

The experiment was performed at the Maier-Leibnitz Laboratory in Garching, Munich, Germany. A 14-MeV proton beam was produced in the tandem accelerator and transported to the target position of the Q3D spectrograph \cite{doi:10.1080/10619127.2018.1427405}. Reaction products were momentum-analysed in the spectrograph and detected in a suite of detectors consisting of two propotional gas counters, one of which is position-sensitive and a plastic scintillator located at the focal plane of the spectrograph \cite{WirthPhD}. Various different targets including $^{12}$C, LiF, SiO$_2$ and NaF were used in the experiment. An example excitation-energy spectrum obtained at a scattering angle of $\theta_\mathrm{Q3D} = 70$ degrees is shown in Fig. \ref{fig:na23pp_spectrum}. 

From the spectrum obtained in the experiment, it is clear that there is no evidence to support the existence of either of these resonances. For this reason, we conclude that the resonances likely do not exist and that they should be omitted from future evaluations of the $^{22}$Ne($p,\gamma$)$^{23}$Na reaction rate.

\section{$^{30}$Si($p,\gamma$)$^{31}$P and the $^{30}$Si($^3$He$,d$)$^{31}$P reaction}

The $^{30}$Si($p,\gamma$)$^{31}$P reaction is a bottleneck when moving material from around magnesium up to potassium. The temperature and density conditions of the polluting site inferred from observed abundance patterns are therefore extremely sensitive to the rate of this reaction. Since the resonance energies of importance for this reaction are so small, the proton widths are typically much smaller than the $\gamma$-ray widths ($\Gamma_p \ll \Gamma_\gamma$ and $\Gamma \approx \Gamma_\gamma$), and the spins of the reactants are $j_\mathrm{Si} = 0$ and $j_p = \frac{1}{2}$ this means that the resonance strength of the reaction simplifies to $\omega\gamma = \frac{2J+1}{2}\Gamma_p$.

For the $^{30}$Si($p,\gamma$)$^{31}$P reaction the missing pieces of information are the spins and parities of the resonance states in $^{31}$P, and the proton partial widths of those states. In order to determine these quantities we used the $^{30}$Si($^3$He$,d$)$^{31}$P one-proton transfer reaction. For these transfer reactions, the shape of the angular distribution is characteristic of the transferred orbital angular momentum, $\ell$, while the overall amplitude of the angular distribution gives information on the spectroscopic factor and the proton partial width.

\begin{figure}
    \centering
    \includegraphics[width=\textwidth]{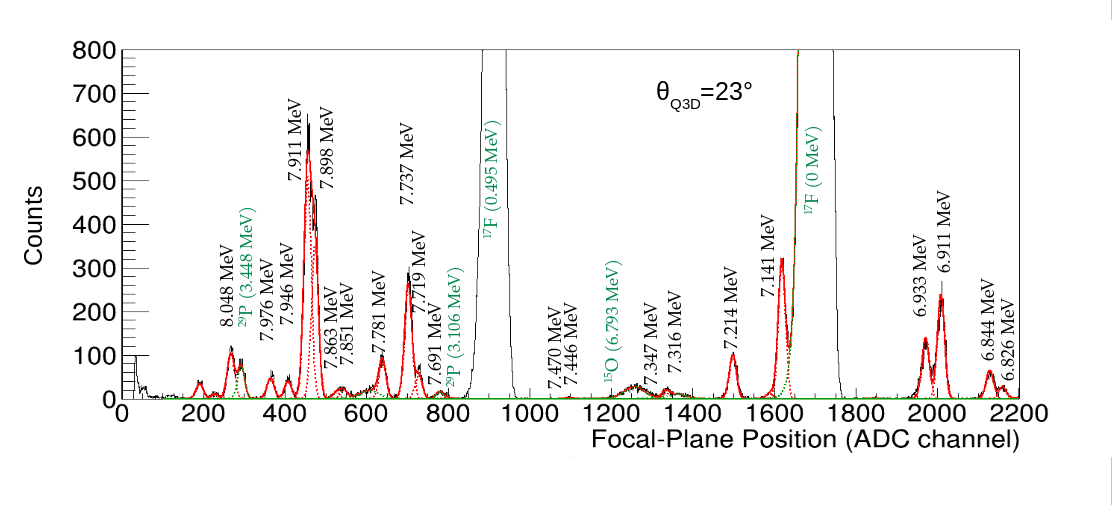}
    \caption{Focal-plane position spectrum for the $^{30}$Si($^3$He$,d$)$^{31}$P reaction taken at $\theta_\mathrm{Q3D} = 23$ degrees. The excitation energies of $^{31}$P states are labelled in black along with the excitation energies and sources of prominent contaminant peaks in green. The fit to the data used to extract the yields is also included.}
    \label{fig:SiSpectrum}
\end{figure}

The experiment was also performed with the Munich Q3D spectrograph. In this experiment the beam was $^3$He at an energy of 25 MeV. Reaction products were momentum-analysed in the Q3D. Deuterons were identified at the focal plane by the differential energy deposition in the focal-plane detectors. An example focal-plane energy spectrum is shown in Fig. \ref{fig:SiSpectrum}.

\begin{figure}
    \centering
    \begin{minipage}[b]{2.5in}
    \caption{Differential cross section for the $^{30}$Si($^3$He$,d$)$^{31}$P reaction to the $E_\mathrm{x} = 7737$-keV state in $^{31}$P. The differential cross section is fitted with the result of an $\ell = 3$ DWBA calculation.}
    \end{minipage}
    \hspace{0.3in}
    \begin{minipage}[b]{3in}
    \includegraphics[width=\textwidth]{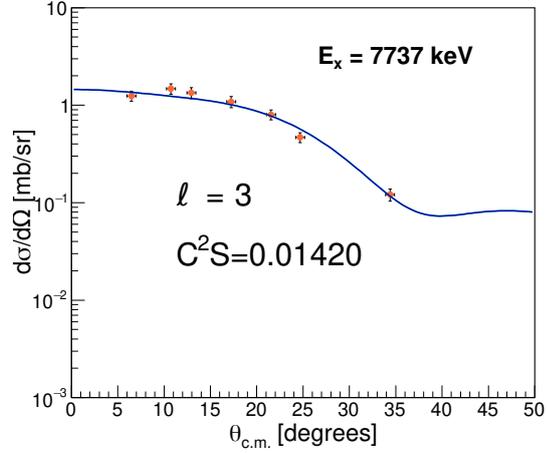}
    \end{minipage}
    \label{fig:DWBA}
\end{figure}

States in $^{31}$P were identified by ensuring that the states were observed at multiple angles with consistent excitation energies. The differential cross sections for the states were extracted and used to compute the proton widths. Differential cross sections were calculated using the code {\sc fresco} assuming single-step sudden transfer reactions under the Distorted-Wave Born Approximation (DWBA). The strength and shape of the differential cross sections could then be used to extract the proton spectroscopic factors and calculate the proton widths, as described in Refs. \cite{PhysRevC.105.015805,ILIADIS1997166}. An example of the data with the associated calculation is shown in Fig. \ref{fig:DWBA}. The wavefunctions used for the calculation of the proton widths were extracted from the DWBA calculations, ensuring consistency between these calculations and reduces some of the uncertainty in the extraction of the proton width, see Ref. \cite{PhysRevC.105.015805} for details.

With these new proton widths, and existing resonance data on $^{30}$Si($p,\gamma$)$^{31}$P, the reaction rate was recomputed using the {\sc ratesMC} code \cite{ILIADIS2010251,RatesMC}. The uncertainties in the reaction rate in the region of interest have been greatly reduced. The major remaining contribution to the uncertainty is in the unknown spin-parity of the $E_r = 149$-keV resonance (the $E_\mathrm{x} = 7446$-keV resonance state). If this state has a proton orbital angular momentum of $\ell = 2$ then it may contribute to the reaction rate but if it is $\ell = 3$ then the contribution is negligibly small. Once this spin-parity is determined then the reaction rate will be well known at all temperatures. For more details as to the experiment and the reaction-rate evaluation, see Ref. \cite{PhysRevC.105.015805}.

\section{A direct measurement of the $^{39}$K($p,\gamma$)$^{40}$Ca reaction with the DRAGON recoil separator}

Direct measurements of resonance strengths are preferred if possible. If, as was done in Ref. \cite{Dermigny_2017}, the astrophysically important resonances have been identified then direct measurements may be targeted so that the resonance strengths or cross sections may be determined. For proton radiative-capture reactions such as those measured at LUNA mentioned above, these experiments may be performed using a proton beam impinging on the target of interest with the resulting $\gamma$-ray decays detected in $\gamma$-ray detectors such as BGO or HPGe.

An alternative approach is to try to detect the heavy recoil produced in the reaction. This requires the experiment to be done in inverse kinematics with a heavy-ion beam incident on a gas target. The heavy recoils must then be separated from the unreacted beam, transported through some ion-optical system and detected at a focal plane.

The DRAGON (the Detector of Recoils And Gamma-rays of Nuclear reactions) at TRIUMF, Vancouver, Canada, is one such device \cite{HUTCHEON2003190}. A differentially pumped windowless hydrogen gas target is surrounded by an array of BGO detectors. Heavy ions, in this case $^{39}$K produced in the Off-Line Ion Source (OLIS) were directed into the target at three different energies, corresponding to the resonances of interest at $E_r = 666$, $606$ and $337$ keV. When a radiative-capture reaction occurred, the resulting $\gamma$ rays were detected in the BGO detectors. The heavy $^{40}$Ca recoils were transported through the recoil separator which consists of a series of dipole, magnetic-electric-magnetic-electric, with various additional focussing elements such as quadrupoles. The heavy recoils were detected in the focal plane in a detector suite consisting of two microchannel plates in front of a silicon detector placed within an ionisation chamber.

\begin{figure}
    \centering
    \begin{minipage}[b]{2.5in}
    \caption{Separator time-of-flight spectrum for the $E_r = 337$-keV resonance of the $^{39}$K($p,\gamma$)$^{40}$Ca reaction. The black spectrum is that which only has gate conditions placed on the silicon detector energy while the red spectrum has additional gates placed on the BGO-RF time difference and the MCP-RF time difference, significantly reducing the background from leaky beam.}
    \end{minipage}
    \hspace{0.2cm}
    \begin{minipage}[b]{3.5in}
    \includegraphics[width=\textwidth]{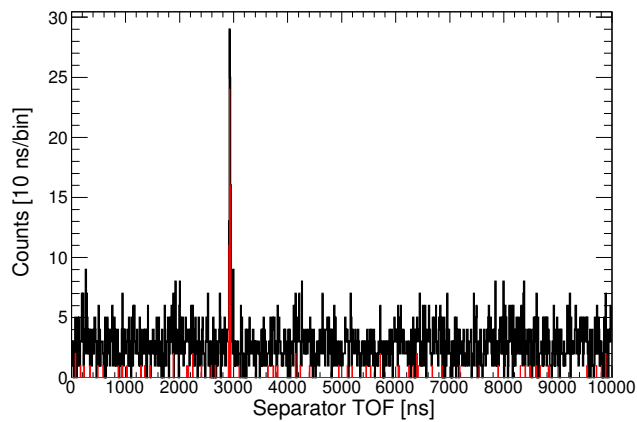}
    \end{minipage}
    \label{fig:tofSpectrum}
\end{figure}

The $^{40}$Ca recoils of interest are identified using various different observables from the current system. Most important is the separator time-of-flight which gives the time elapsed between a $\gamma$-ray being detected in the BGO detectors and the heavy recoil being detected in the focal plane. Other useful observables in reducing the background from so-called \textquoteleft leaky beam', $^{39}$K beam ions which manage to hit the focal plane through multiple scattering within the separator, including the time difference between the accelerator RF signal and the BGO hit which gives rudimentary information about where in the target the reaction took place. An example separator time-of-flight spectrum is shown in Fig. \ref{fig:tofSpectrum}

For these DRAGON data, the beam normalisation and yield extraction are complete. The only outstanding parts of the analysis are accounting for the overall detector efficiency by taking into account the $\gamma$-ray decays from the compound $^{40}$Ca resonances, and the charge-state distribution of the outgoing $^{40}$Ca recoils. The observed $\gamma$-ray intensities following the proton-capture reaction do not follow the expected pattern using the listed intensities from a previous study of this reaction \cite{KIKSTRA1990425}. A pending experiment at iThemba LABS using the AFRODITE array of HPGe clovers will provide complementary information which will help to resolve some of the these discrepancies.

\section{Concluding remarks and summary}

The evolution of globular cluster is key to understanding how galaxies have merged and grown but their history is complicated by multiple stellar populations which pollute these objects. Due to the variety of problems in the nuclear data, ranging from whether states exist through the energies, and spins and parities of states to the partial widths and resonance strengths, a number of different experimental techniques are required. This paper highlights three different experiments and discusses how the various nuclear data obtained from them can be used to better understand the history of globular clusters.

\bibliographystyle{iopart-num}
\bibliography{refs}


\end{document}